\documentclass[apjl]{emulateapj}

\usepackage{amsmath}
\usepackage{graphicx}
\usepackage{epsfig,psfrag,epic,eepic}
\usepackage{textcomp}
\usepackage{times}

\slugcomment{Accepted for publication in ApJ Letters, April 7, 2011}

\shorttitle{Piercing the Glare: Direct Imaging Search for Planets in the Sirius System}
\shortauthors{Thalmann et al.}

\newcommand{\Bra}{\textrm{Br}\,\ensuremath{\alpha}}

\begin{document}

\title{Piercing the Glare: A Direct Imaging Search for Planets in the Sirius System\altaffilmark{$\star$}}

\author{C. Thalmann\altaffilmark{1,2}, 
	T. Usuda\altaffilmark{3}, 
	M. Kenworthy\altaffilmark{4},
	M. Janson\altaffilmark{5}, 
	E.~E. Mamajek\altaffilmark{6},
	W. Brandner\altaffilmark{2},
	C. Dominik\altaffilmark{1,7},
	M. Goto\altaffilmark{2},
	Y. Hayano\altaffilmark{3},
	T. Henning\altaffilmark{2},
	P.~M. Hinz\altaffilmark{8},
	Y. Minowa\altaffilmark{3},
	M. Tamura\altaffilmark{9}
}

\altaffiltext{$\star$}{Based on data collected at Subaru Telescope, which
	is operated by the National Astronomical Observatory of Japan, and at
	the MMT Observatory, a joint facility of the University of Arizona and
	the Smithsonian Institution.
}
\altaffiltext{1}{Anton Pannekoek Astronomical Institute, University of Amsterdam,
	The Netherlands; \texttt{thalmann@uva.nl}.} 	
\altaffiltext{2}{Max Planck Institute for Astronomy, Heidelberg, Germany.}
\altaffiltext{3}{Subaru Telescope, Hilo, Hawai`i, USA.}
\altaffiltext{4}{Leiden Observatory, Leiden University, P.O. Box 9513, 2300 RA Leiden,
The Netherlands.}
\altaffiltext{5}{University of Toronto, Toronto, Canada.}
\altaffiltext{6}{University of Rochester, Rochester NY, USA.}
\altaffiltext{7}{Radboud University, Nijmegen, The Netherlands.}
\altaffiltext{8}{University of Arizona, Tucson AZ, USA.}
\altaffiltext{9}{National Astronomical Observatory of Japan, Tokyo, Japan.}

\begin{abstract}\noindent
Astrometric monitoring of the Sirius binary system over the past century
has yielded several predictions for an unseen third system component, 
the most recent one suggesting a $\lesssim$50\,$M_\textrm{Jup}$ object in a
$\sim$6.3-year orbit
around Sirius~A.  Here we present two epochs of high-contrast imaging observations 
performed with Subaru IRCS and AO188 in the 4.05\,$\mu$m narrow-band \Bra{}  
filter.  These data surpass 
previous observations by an order of magnitude in detectable companion
mass, allowing us to probe the relevant separation 
range down to the planetary mass regime (6--12\,$M_\textrm{Jup}$ 
at 1\arcsec, 2--4\,$M_\textrm{Jup}$ at 2\arcsec, and 1.6\,$M_\textrm{Jup}$ 
beyond 4\arcsec).  We complement these data with one epoch of
$M$-band observations 
from MMT/AO Clio, which reach comparable performance.  No dataset  
reveals any
companion candidates above the 5\,$\sigma$ level, allowing us to refute 
the existence of Sirius C as suggested by the previous astrometric 
analysis.  
Furthermore, our \Bra{} photometry of Sirius B confirms the lack of an 
infrared excess beyond the white dwarf's blackbody spectrum.
\end{abstract}


\keywords{planetary systems --- techniques: high angular resolution --- 
stars: individual (Sirius A) --- stars: individual (Sirius B) ---
white dwarfs}




\section{Introduction}

One and a half centuries ago, Sirius was found to host a faint binary 
companion  \citep{bond1862}.  The orbital motion of this 
pair has been monitored ever since, leading to a number of publications
that claimed to find periodic perturbations indicative of the presence
of an unseen third system component, Sirius C.  The most recent 
analysis predicted a substellar companion in a $\sim$6.3-year 
circumstellar orbit around Sirius A \citep[and references 
therein]{benest95}.  While the amplitude of the purported astrometric 
signal, 56\,mas, would suggest a companion mass of 72\,$M_\textrm{Jup}$,
these authors imposed an upper limit of $\lesssim$\,50\,$M_\textrm{Jup}$
on the basis of system stability considerations.  Since no measure of
confidence is given, we assume a conservative lower limit of half the 
measured amplitude, 28\,mas, on the basis of their plotted periodograms,
resulting in a minimal mass of 36\,$M_\textrm{Jup}$.

This would place Sirius C in the so-called brown 
dwarf desert, the range of orbital parameter space around stars in which
brown dwarfs are found to be scarce \citep[e.g.][]{marcy00, grether06}.
Numerical simulations of the formation and evolution of brown dwarf
companions reproduce this scarcity, regardless of the formation process 
assumed (e.g.\ as part of the star formation process \citep{bate05},
planet formation by core accretion \citep{mordasini09} or 
gravitational fragmentation \citep{stamatellos09}), thus these rare
objects impose important constraints on theory.

However, Sirius' extreme brightness -- both apparent and absolute --
had long thwarted attempts to verify 
these claims through direct imaging.  \Citet{KB00} established
first constraints on substellar companions at separations of 
1\farcs5--3\arcsec{}
with space-based coronography at 1.02\,$\mu$m on HST NICMOS, whereas
\citet{BBP08} used ground-based observations assisted by adaptive optics
on ESO ADONIS to achieve similar constraints in the range of 
3\arcsec--10\arcsec.  These limits left most of the parameter
space in which Sirius C was expected unexplored.  In this work, 
we present new results from 4.05\,$\mu$m observations on Subaru IRCS
as well as from $M$-band observations on MMT/AO Clio, 
which improve the companion mass constraints by an order of
magnitude and extend the coverage down to an inner working angle of
0\farcs7.  With a baseline of 4.3 years among the observations,
the chance that a companion be missed in all datasets due to adverse 
geometric conditions is slim.

Sirius A is an A1V-type star at a distance of 2.64\,pc, a mass of 
2.02\,$M_\odot$ and an age of
225--250\,Myr, whereas Sirius B is a white dwarf of
0.98\,$M_\odot$ with a cooling age of 124\,$\pm$\,10\,Myr orbiting Sirius 
A with a 50-year period \citep[e.g.][]{liebert05}.  The combination of 
youth and extreme proximity \citep[cf.\ a median target 
distance of 22\,pc in the
Gemini Deep planet survey,][]{geminideep} allow us to explore
unusually small orbital radii, with planetary-mass detection limits down 
to separations of 2.5\,AU in projection.

\section{Observations}

\begin{figure*}[p]
\centering
\includegraphics[width=0.92\linewidth]{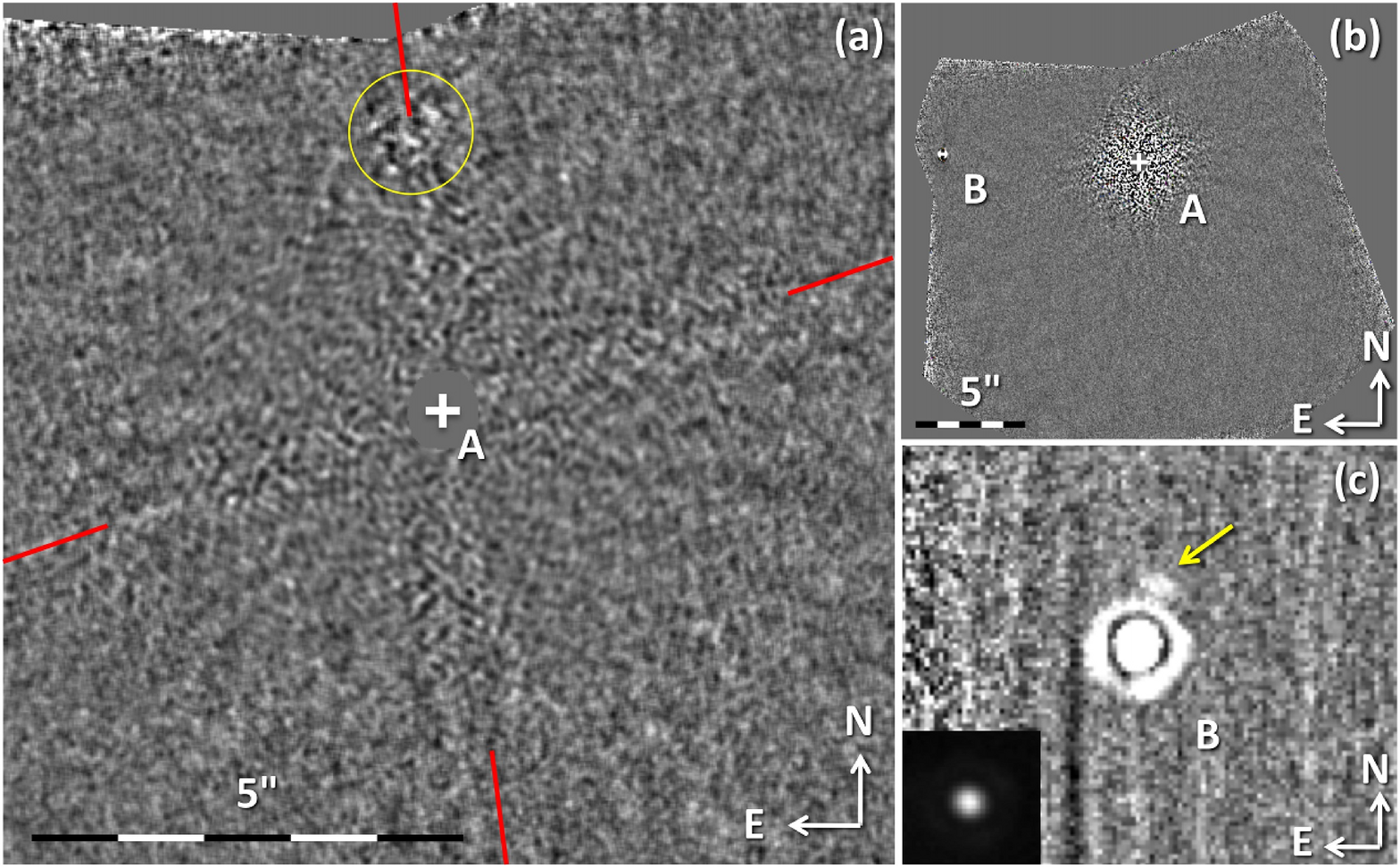}
\includegraphics[width=0.92\linewidth]{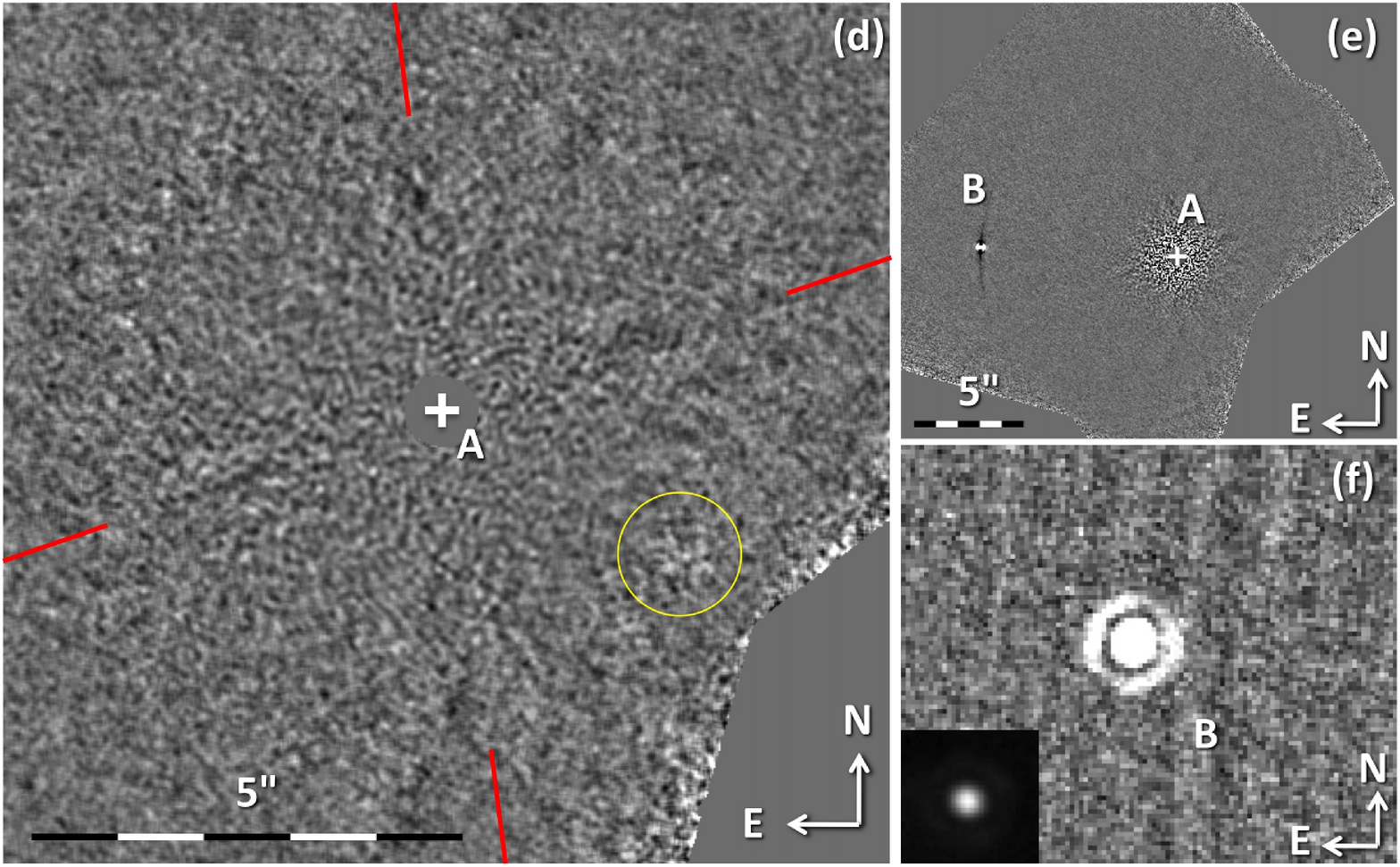}
\caption{(a--c) Results of the January 2011 ADI observations of the 
	Sirius system in the \Bra{} band with Subaru IRCS.
	(a) Signal-to-noise map of the inner $\sim$10\arcsec$\times$10\arcsec, 
	calculated in concentric annuli 
	after convolution with a 4\,px diameter circular aperture.  
	The stretch is $[-5\,\sigma,5\,\sigma]$. The 
	location of Sirius A is marked with a white plus sign. The short
	red lines indicate the position angles where residuals of the 
	spider diffraction pattern slightly elevate the noise level.  The
	yellow circle marks the area of increased noise generated by one of
	the unused mirror holes in the field of view.
	(b) Full field of view of the ADI intensity 
	image. Sirius B is visible to the East of Sirius A. (c) 
	Non-differential image of Sirius B. Unsharp masking on the scale of
	15 pixels (3\,$\lambda/D$) has been applied to reduce background 
	structures.  The point source just outside Sirius B's first Airy
	ring is a ghost caused by a filter.  The cutout shows Sirius
	B's unsaturated PSF core. (d--f) The corresponding images for the
	March 2011 Subaru IRCS data.
	\vspace*{3mm}
}
\label{f:bra}
\end{figure*}

The Sirius system was observed with the Subaru IRCS instrument
\citep{IRCS} on January 20, 2011,
using the AO188 adaptive optics system \citep{AO188}.  The \Bra{} 
narrow-band filter at 4.05\,$\mu$m was chosen, since this band has 
been shown both theoretically and observationally to provide
the best combination of PSF quality and brightness contrast
for planet detection in the speckle-dominated regime \citep{janson08, 
janson09, janson10}.  The dataset consists of 56 frames, each 
comprising 240 co-adds of 0.1\,s exposures, for a total integration 
time of 22.4\,min.
Although IRCS does not offer Lyot coronography, a mirror plate with
several circular holes is available to reject light from the primary 
star.  We used a hole with a projected radius of 0\farcs3 to avoid
excessive saturation; nevertheless, the shoulders of 
Sirius A's point-spread function (PSF) locally saturate out to a radius
of $\sim$0\farcs6.  The plate scale was 20\,mas per pixel.

Follow-up observations were taken with Subaru IRCS on March 12, 2011.
Using the same observing strategy, 83 frames were taken for 33.2\,min
of total integration time.

Furthermore, we make use of $M$-band imaging data taken on December 5, 2006, 
with the Clio 3--5$\,\mu$m imager \citep{freed04} in conjunction with 
the adaptive secondary mirror on the 6.5\,m MMTO telescope 
\citep{brusa04}. 
The star was nodded 5\farcs5 along the
long axis of the detector after five images were taken. Each of
the 322 images consists of 50 co-added exposures of 209.1 ms
length, for a total co-added duration of 56.1 min. To avoid
variations in the pattern of illumination on the Clio detector, the
instrument is fixed in orientation with respect to the telescope,
resulting in total field rotation of 24.6$^\circ$ for our Sirius data.
Conditions were photometric, and the seeing 0\farcs5--0\farcs7 
(as seen with the video rate optical acquisition camera) at an air 
mass of 1.50--1.64.  The plate scale was 49\,mas per pixel.

All observations were taken in pupil-stabilized image orientation
to enable the 
angular differential imaging technique \citep[ADI,][]{marois06}.  We
employed the LOCI algorithm \citep[Locally Optimized Combination
of Images,][]{lafreniere07} in order to search for faint point
sources in Sirius A's speckle halo.  This form of ADI is the most 
powerful high-contrast imaging method currently available, as 
evidenced by recent direct detections of substellar companions 
\citep[e.g.][]{thalmann09, marois10} and even circumstellar
disks \citep{thalmann10, buenzli10}.  We employed the LOCI 
parameters described as optimal for the test dataset in 
\citet{lafreniere07}, with a 
frame selection criterion of 0.5\,FWHM to avoid self-subtraction
of companion signals.  Given the excellent resulting image quality,
we refrained from further parameter fine-tuning.  In all datasets, 
the inner working angle of 0\farcs7 is due to detector nonlinearity
and saturation rather than insufficient field rotation.

\section{Results}

\subsection{Subaru IRCS \protect{\Bra{}} Data}

The final \Bra{} images after ADI reduction are presented in 
Figures~\ref{f:bra}a--f.  No obvious point-sources
around Sirius A
are visible.  To confirm this numerically, we first convolve the
image with a sampling 
aperture of 4 pixel diameter, and then calculate the
signal-to-noise map (S/N) by dividing the pixel values in 
concentric 
annuli around the star by their standard deviation.  As expected, 
we find no
signal above the $5\,\sigma$ level, discounting the 
locally elevated noise in the shadow of the unused 
mirror holes.

We do find a conspicuous signal in the immediate vicinity of Sirius B
in the January 2011 data (Figure~\ref{f:bra}c). However, closer 
inspection of the signal's behavior
in the time-series of images reveals it to be locked to the position
angle of the pupil rather than the field, and identifies it as a
ghost.  No such signal is seen in the March 2011 data.  Apart from
this, the environs of Sirius B are background-limited outside the 
first Airy ring.

We use the unsaturated PSF of Sirius B in the \Bra{} images for
photometric calibration.  Although no prior \Bra{} photometry of 
Sirius B exists, its overall brightness in the near- to mid-infrared
can be predicted by a 25,193\,$\pm$\,37\,K blackbody 
\citep{barstow05,skemer11}.   Model 
spectra by \citet{lejeune97} for a range of $\log g$ values up to 
5\,cm/s$^2$ for $T_\textrm{eff} =$ 25,000\,K indicate that while 
absorption lines are present in the mid-infrared, they exhibit rather
small equivalent width. For the \Bra{} absorption line we estimate an
equivalent width of 0.7\,nm, and hence conclude that it does not 
present a significant deviation from a blackbody.

As an independent 
confirmation, we calibrate Sirius B's brightness using a brief 
observation of the A1V-type star HD~40138 as a reference.  The 
result, 9.27\,$\pm$\,0.17\,mag, is consistent with our expected 
value of 9.17\,$\pm$\,0.10\,mag, which is based on the $J$-band photometry
by \citet{BBP08} and stellar spectra by \citet{castelli03}.
This provides further evidence that Sirius B
does not have an infrared excess from circumstellar dust as 
proposed by \citet{BBP08}. \looseness=-1

\subsection{MMT/AO Clio \protect{$M$}-band data}

\begin{figure}[t]
\centering
\vspace{2mm}
\includegraphics[width=\linewidth]{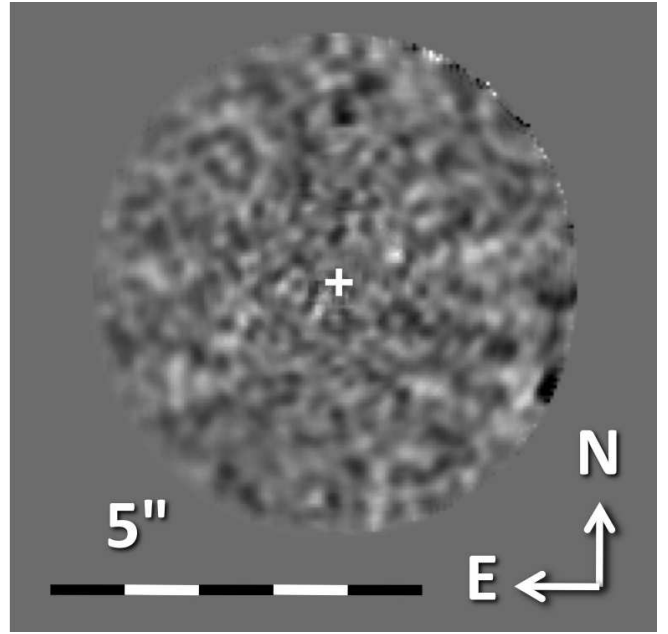}
\caption{Signal-to-noise map of the inner 4\arcsec$\times$4\arcsec of 
	the MMT/AO Clio $M$-band
	data.  The stretch is $[-5\,\sigma,5\,\sigma]$.
	No significant signal is found.	
	}
\label{f:mband}
\vspace*{3mm}
\end{figure}

Individual images from Clio are beam-subtracted, and a sigma clipping
routine is used to remove hot pixels deviating by more than $3\,\sigma$
from a 5\,$\times$\,5 pixel region centered on that pixel. The 
location of the
central star is estimated by smoothing the images with a Gaussian
kernel with FWHM 10 pixels and estimating the location of the
resultant peak. All the science images are then passed into a LOCI
data reduction pipeline \citep{lafreniere07}. The LOCI processing
carries out reduction in 4 pixel (0\farcs2) wide rings starting
from an inner radius of 4 pixels (0\farcs2). Each optimization
section extends 20 pixels beyond the outer radius of the current ring,
ensuring that the optimization section contains the equivalent area of
a minimum of 150 PSF cores. After the LOCI background subtraction is
performed, the resultant images are rotated into the frame of the 
sky before being averaged together to give a final output image.

Again, no significant signal above the 5\,$\sigma$ threshold is found.
The S/N map is shown in Figure~\ref{f:mband}.

\subsection{Contrast and detectable companion mass}

\begin{figure}[t]
\centering
\includegraphics[width=\linewidth]{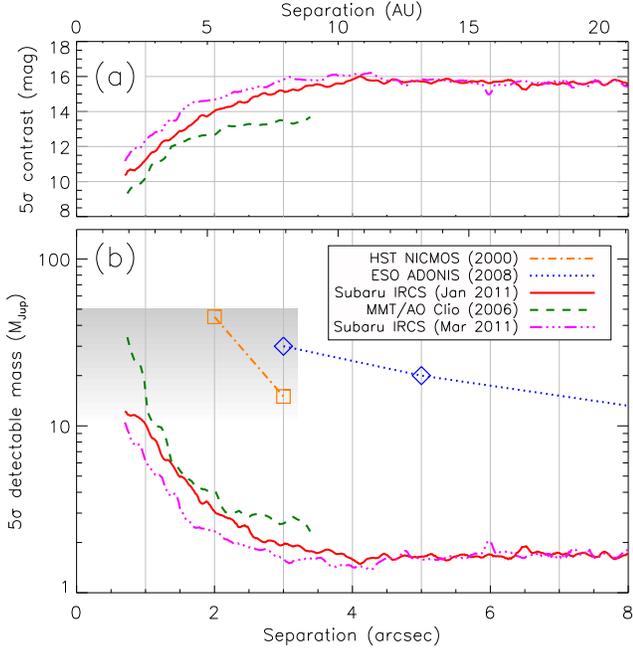}
\caption{Limits on faint companions to Sirius A. (a) Effective
	$5\,\sigma$ detectable contrast to Sirius A in the January
	2011 \Bra{} data, (solid red), the $M$-band data (dashed green),
	and the March 2011 \Bra{} data (dash-dot-dot-dotted magenta)
	as a function of separation, corrected for partial 
	subtraction. (b) The companion masses corresponding to those
	contrasts, derived according to an updated version of the 
	method of \protect{\citet{janson08}} from the \texttt{COND}
	evolutionary model \protect{\citep{allard01,baraffe03}}.  For 
	comparison, the constraints
	established by \protect{\citet[dash-dotted orange]{KB00}} and 
	\protect{\citet[dotted blue]{BBP08}} are overplotted.  The gray shaded
	area indicates the parameter space of a 
	$\lesssim$50\,$M_\textrm{Jup}$ companion in a 6.3-year orbit
	(semi-major axis $a=1\farcs6$).
	}
\label{f:contrast}
\vspace*{1mm}
\end{figure}

The $5\,\sigma$ contrast curves established by the three datasets is 
presented in Figure~\ref{f:contrast}.  The partial self-subtraction
of point sources during the LOCI reduction is estimated by inserting 
artificial 10\,$\sigma$ point sources into the raw data across the 
entire usable range of separations from the star, and 
retrieving their surviving flux after subjecting them to the LOCI 
pipeline (radial filtering and optimized background subtraction).
The plotted curves are corrected for this flux loss and therefore 
represent effective detectable contrast.  The process is described 
in detail in \citet[Section 4.3.]{lafreniere07}.  

The \texttt{COND}-based 
evolutionary models \citep{allard01,baraffe03} are used to convert
the $M$-band
brightness contrast into detectable mass.  For the \Bra{} data,
we adopt an updated version of the method used in 
\citet{janson08}. \texttt{COND} model spectra \citep{allard01} are 
used to re-calculate photometric predictions from \texttt{COND}-based 
evolutionary models \citep{baraffe03} from $L^\prime$ band to the IRCS 
\Bra{} band, for masses of 1--20\,$M_\textrm{Jup}$ and the age of the 
Sirius system (250\,Myr). The measured brightness
contrast as function of angular separation is then translated into
detectable mass, using these values. 

The \Bra{} data are sensitive to masses in the 
planetary regime down to the inner working angle of 0\farcs7 around
Sirius A (6--10\,$M_\textrm{Jup}$ at 1\arcsec, 2--3\,$M_\textrm{Jup}$ at 
2\arcsec), and reach 1.6\,$M_\textrm{Jup}$ in the thermal background
beyond 4\arcsec, which includes the immediate surroundings of 
Sirius B down to 0\farcs2.  The $M$-band data achieve comparable 
performance in a more restricted field of view ($<$3\farcs5).

\section{Discussion}

Our high sensitivity to substellar companions around Sirius A allows 
the first thorough test of the hypothesis of a $\lesssim$$50\,M_\textrm{Jup}$ 
body, ``Sirius C'', in a 6.3-year orbit around Sirius A as proposed by 
\citet{benest95}. 
As these authors point out, though, a single negative 
detection does
not prove the non-existence of such an object.  Even though the 
semi-major axis of such an orbit is $a=4.3\,\mathrm{AU}$, 
corresponding to 1\farcs6 at the distance of 2.64\,pc, both projection 
effects and eccentricity can leave the companion at an apparent 
separation below the inner working angle of $0\farcs7$ at the time of
observation. \looseness=-1

To quantify the likelihood of a false negative result, we 
systematically calculate the projected orbital ellipses for 6.3-year
orbits with eccentricities $e = \{0, 0.04, \ldots, 0.96\}$ and arguments
of periastron $\omega = \{0\degr, 10\degr, \ldots, 350\degr\}$.  Since 
we make use of radial detectable mass curves (Fig.~\ref{f:contrast}), 
we are insensitive to the orbit's overall position angle on sky; thus no
variation of the longitude of the ascending node $\Omega$ is necessary.
As for the inclination, we explore two scenarios: The coplanar case, 
where Sirius C shares the orbital plane of the Sirius AB system with
an inclination of 136.6$^\circ$, and the free case, where all system 
orientations on the unit sphere are considered equally likely, 
resulting in a statistical weight of $\sin i$ for each inclination
$i = \{0\degr, 1\degr, \ldots, 90\degr\}$.  The mass of Sirius C is 
sampled as $m = \{0.5, 1, \ldots, 20\}\,M_\textrm{Jup}$.  

\begin{figure}[tb]
\centering
\includegraphics[width=\linewidth]{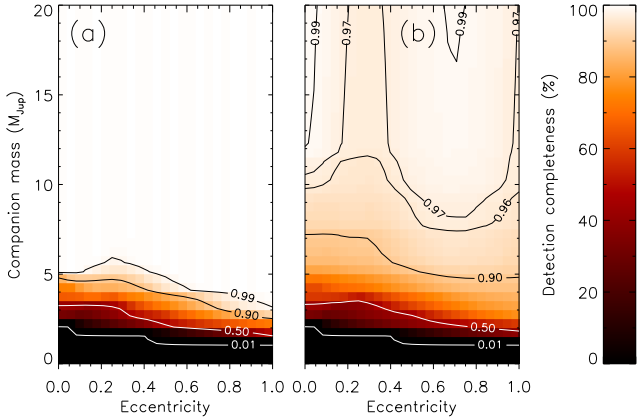}
\caption{Probability of at least one $5\,\sigma$ detection of a 
companion with a 6.3-year orbit around Sirius A in our three datasets,
as a function of companion mass $m$ and eccentricity $e$. (a) 
The case of coplanar orbits with the 
Sirius AB system (inclination $i=136.6^\circ$).  All companions above
6$\,M_\textrm{Jup}$ can be rejected regardless of orbital phase 
(100\% completeness). The
completeness remains above 50\% down to 2--3.5\,$M_\textrm{Jup}$ depending
on eccentricity.  
(b) If the inclination is unconstrained ($p(i) \propto \sin i$), edge-on orbits
become possible, introducing a chance of a few percent that Sirius C might
elude detection in all observations.
Nevertheless, our three datasets provide decisive evidence against the 
existence of Sirius C as proposed by \protect{\citet{benest95}}.
}
\label{f:completeness}
\end{figure}

On each ellipse, 1000 possible companion positions at the
epoch of our $M$-band observations are chosen, with uniform spacing
in eccentric anomaly.  We then calculate the
corresponding location at the epoch of the other two datasets.  
The detectable mass curves are evaluated to determine 
whether or not the companion would have been detected at 
$\ge$$5\,\sigma$ at least once in that configuration.
For each ellipse, the detection likelihood is averaged over all 
companion positions, whereby a statistical weight is
assigned to each position to account for the non-uniform evolution 
of eccentric anomaly with time.  The weight is proportional to the 
time that the companion spends at each position over the course of 
an orbit, approximated by half the distance to the two 
adjacent positions in the sequence divided by the local orbital 
velocity.  Finally, the
detection likelihoods are averaged over all $i$ and $\omega$ to yield
the detection completeness for each combination of eccentricity $e$ 
and companion mass $m$.  The results of this analysis are plotted in 
Figure~\ref{f:completeness}.

\begin{figure}[t]
\centering
\includegraphics[width=\linewidth]{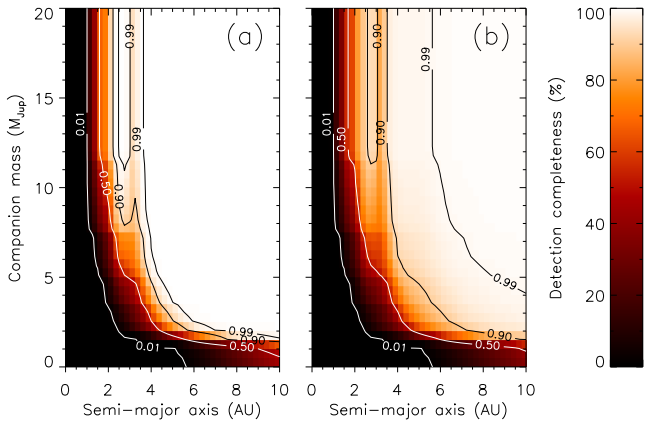}
\caption{Probability of at least one $5\,\sigma$ detection of a 
companion around Sirius A in our three datasets,
as a function of companion mass $m$ and semi-major axis $a$.  A flat
distribution in eccentricities is assumed.  (a) 
The case of coplanar orbits with the 
Sirius AB system (inclination $i=136.6^\circ$).  (b) The case of 
unconstrained inclination ($p(i) \propto \sin i$).  The false negative
probability is large for the stable planet orbit regime at
$a \lesssim 2\farcs17$ \protect{\citep{holman99}}.
}
\label{f:completeness_planets}
\end{figure}

We find that coplanar companions down to 6\,$M_\textrm{Jup}$ can be 
excluded at the $5\,\sigma$ level at 100\% completeness.  The 
completeness remains above 50\% down to 2--3.5\,$M_\textrm{Jup}$, 
depending on eccentricity.
If no constraints are imposed on the inclination, edge-on orbits 
emerge that can hide Sirius C behind Sirius A's glare.  As a result,
the
completeness values for masses above 12\,$M_\textrm{Jup}$ drop to 
97--99\% for certain eccentricity ranges, and down to 90\%
for masses of 5--7\,$M_\textrm{Jup}$.

The astrometric signal reported by \citet{benest95}, 56\,mas, implies
a companion mass of 72\,$M_\textrm{Jup}$.  
The authors furthermore impose an upper mass limit of 
$\lesssim$50\,$M_\textrm{Jup}$ on the basis of system stability 
considerations.  While no lower mass limit is given, we derive a 
conservative estimate of half that value, 
36\,$M_\textrm{Jup}$, from their published periodograms.  
Given our much lower detection limits, our three 
combined epochs of high-contrast imaging can therefore decisively 
reject their Sirius C hypothesis.  

Although \citet{benest95} provide no error estimation, we consider a
false alarm the most likely explanation for their results.
Precision astrometry is known to suffer from 
a multitude of systematic errors (e.g.\ subtle changes in pixel scale
and orientation, and differential atmospheric refraction which depends
on airmass, parallactic angle, and ambient conditions), and has led
to a series of spurious detections in the past (e.g. \citet{pravdo09}
vs.\ \citet{bean10} and \citet{lazorenko11}; \citet{vandekamp69} vs.\
\citet{gatewood73}).

Leaving astrometric predictions aside, we can also explore the 
parameter space for other semi-major axes.  
Figure~\ref{f:completeness_planets} shows the completeness as a
function of semi-major axis $a$ = $\{0.25,$ $0.50,$ $\ldots,$ $10.0\}$\,AU 
and companion mass $m$, assuming a
flat distribution in eccentricity.  Although our data
are sensitive to planets down to an inner working angle of $0\farcs7$
and down to 1.6\,$M_\textrm{Jup}$ at large separations, the 
completeness values drop quickly at shorter separations (e.g.\ 50\%
for a 10\,$M_\textrm{Jup}$ object at $a$ = 2\,AU).  This coincides with the 
domain of long-term stable planet orbits predicted by \citet{holman99},
with a critical semi-major axis $a_\textrm{c}=2.17$\,AU.  Therefore, 
plenty of parameter space remains for unseen planets around Sirius A.  
The 
upcoming next generation of high-contrast instruments, such as SPHERE
\citep{beuzit10}, will offer smaller inner working angles and better
contrast performance, and thus stand a good chance to detect such
planets.  In particular, due to 
its extreme proximity and brightness, Sirius is the third most 
promising target (after $\alpha$ Centauri A and B) for the direct 
detection of exoplanets in reflected light with the SPHERE ZIMPOL
imaging polarimeter \citep{thalmann08}. \looseness=-1

One thing to keep in mind is the fact that Sirius B was originally a
$\sim$5\,$M_\odot$ progenitor star that expanded into a supergiant 
$\sim$125\,Myr ago, with potentially dramatic consequences for the
system architecture \citep{liebert05}.  Accretion of ejected material
from Sirius B may have caused planets around Sirius A to gain mass and
heat, migrate, or form in the first place as second-generation planets
\citep[e.g.][]{perets10}.
Since these processes would leave the planets hotter and brighter than
their unperturbed 250\,Myr-old counterparts, our detectable planet
mass curves in Figure~\ref{f:contrast} are conservative for such
objects.

\section{Conclusions}


We present three high-contrast imaging datasets of the Sirius system,
all of which reach detection performances in
the planetary regime (6--12\,$M_\textrm{Jup}$ at 1\arcsec, 
2--4\,$M_\textrm{Jup}$ at 2\arcsec, 1.6\,$M_\textrm{Jup}$ beyond
4\arcsec).  This constitutes an improvement of an
order of magnitude in detectable planet mass.  Taken up to 4.3 years 
apart,
the observations allow us to refute the existence of a substellar
companion with a mass of $\lesssim$50\,$M_\textrm{Jup}$ in a 6.3-year
orbit as predicted from astrometry measurements of the Sirius AB
system \citep{benest95}.  For a companion mass above 
12\,$M_\textrm{Jup}$, the chances of a triple false negative at a
5\,$\sigma$ threshold are 0--4\%, depending on eccentricity.  For the
special case of coplanar
orbits, the probability is 0\% down to 6\,$M_\textrm{Jup}$).  
However, we note that our observations leave open the possibility
for Jupiter- and Neptune-sized planets around Sirius A, especially at
short angular separations.

Furthermore, we confirm the absence dust
around Sirius B by the lack of an infrared excess at 4.05\,$\mu$m
within our precision of 0.17\,mag.


\acknowledgements
We thank David Lafreni\`ere for generously having provided us with the
source code for his LOCI algorithm.  We are grateful for the privilege 
of observing from Mauna Kea, which holds great cultural significance for
the Hawai`ian indigenous community.

{\it Facilities:} \facility{Subaru (IRCS, AO188)}, \facility{MMT/AO 
(Clio)}.

\clearpage

\end{document}